# Measuring a Six-hole Recorder Flute's Response to Breath Pressure Variations and Fitting a Model

Daniel Chin
NYU Shanghai
daniel.chin@nyu.edu

Gus Xia
NYU Shanghai
gxia@nyu.edu

## ABSTRACT

We propose the *Siamese-flute* method that measures the breath pressure and the acoustic sound in parallel. We fit a 6-DoF model to describe how the breath pressure affects the octave and the microtonal pitch bend, revealing the *octave hysteresis*. We release both our model parameters and our data analysis tools.

## 1. Introduction

The plastic six-hole recorder flute ("**flute**") is played by A) using fingers to cover its key holes, and B) blowing into its mouthpiece. The fingering determines the *pitch class[1]* of the produced sound. The breath determines the *octave* and the microtonal *pitch bend*.

However, the (breath) pressure threshold for octaves depends on the pitch class and demonstrates hysteresis. The way the pitch bend responds to pressure is non-linear and also depends on the pitch class. In this paper, we aim to create a simple model that takes the pitch class and the pressure as input and gives the octave, the pitch bend, and the amplitude as output. We succeed in modeling the octave and the pitch bend and fail to model the amplitude.

In section 2 we describe how we measure the pressure and the acoustic sound in parallel. Section 3 demonstrates how we prepare the data. In section 4 we present the modeling process and results.

### 1.1. Research context

This paper shares the results of a follow-up study after [1]. In [1] I used a simplified model that assumed many things to be linear. After the submission of [1], we conducted studies aiming to develop that model into a still-simple yet much more accurate one.

## 2. Measuring: the Siamese-flute method

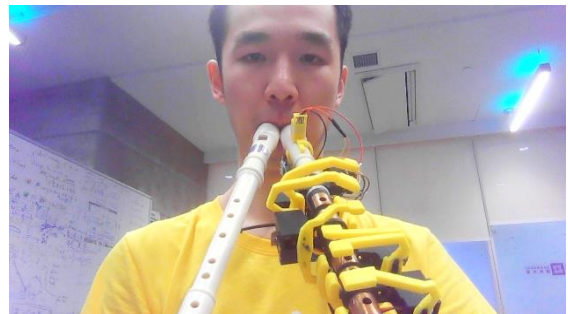

*Figure 1. Siamese Flute.*

The Siamese-flute method requires the player to blow into two flutes simultaneously. In Figure 1, the left flute produces sounds acoustically, while the right flute measures the (breath) pressure using a `BMP085` sensor. Because of the symmetrical placement, we assume the air pressure inside both flutes' mouthpieces to be equal.

The pressure sensor readings, time-coded, are transmitted to the laptop via Serial communication. The Arduino script (`./ardu/ardu.ino`) and the Python data logger (`./logger.py`) is on our GitHub repo[2].

After the recording begins, the player first plays several fast impulses for synchronization purposes (see section 3.3). The player then iterates through seven fingerings, sometimes using tape to seal some key holes in order to free fingers. For each fingering, the player gradually blows harder, reaches a maximal breath velocity, then gradually blows softer. Each fingering is repeated four times.

At this point, we have parallel data: `./test1.mp3` for audio and `./test1.csv` for pressure.

## 3. Preparing the data

We use a Jupyter Notebook (`./lab.ipynb`) to prepare the data.

---

[1] For example, in "C4", "C" is the *pitch class*, and "4" is the *octave*.
[2] 

### 3.1. Resample the pressure array

While the audio signal has a static sample rate, the pressure readings are collected as fast as possible, so the sample rate fluctuates as the Arduino CPU gets random interrupts. This is why we have to time-code the pressure measurements. With the time codes, we generate a static-rate pressure array using linear resampling.

We resample the audio signal into 22050 samples per second and analyze it with a hop length of 512 samples. Therefore, there are about 43.1 hops per second. For convenience, the sample rate of the pressure array is set to equal to the hop rate of the audio signal.

### 3.2. Estimate frequency and amplitude

To estimate the fundamental frequency (f0) of a given audio page (2048 samples), we use the yin algorithm [2]. To estimate the amplitude, we sum the energy in all frequency bins given by the periodogram of the audio page, and divide it by the page size.

### 3.3. Time alignment / synchronization

We use the two groups of five fast impulses played at the beginning of the session to determine the offset between the pressure array and the audio signal.

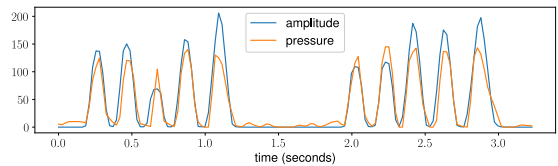

*Figure 2. Alignment, beginning.*

Figure 2 shows good alignment, limiting timing error below one hop. Then, we check the last two notes of the session:

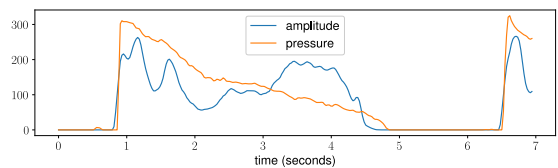

*Figure 3. Alignment, ending.*
*The x axis "0" is ad-hoc.*

Figure 3 shows good attack alignment, which means the Arduino clock and the audio recorder (a laptop, in our case) clock did not drift apart.

### 3.4. Remove disequilibrium near octave jumps

After successful alignment, we plot the pressure, pitch, and amplitude against time in Figure 13 (upper half). To better make sense of Figure 13, you may want to view it while listening to `./test1.mp3`. Visual inspection at this point can already identify:

- The pitch bend rises as the pressure increases.
- The octave threshold rises as the pitch rises.
- Octave hysteresis. The pressure curve height is usually higher when the pitch curve jumps up than when the pitch curve jumps down.
- Octave jumps seem to disrupt the fluid flow in the flute. See Figure 4.

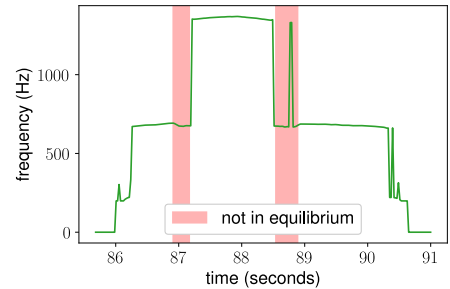

*Figure 4. Disequilibrium around octave jumps.*

The disruption affects the produced pitch. It takes about 300 milliseconds for the fluid flow to restore its equilibrium. We do not study this effect, and simply discard the audio pages around octave jumps. As a convenient coincidence, our algorithm happens to also discards all silent audio pages. The overall removal of data is shown by the pink areas in Figure 13 (lower half).

## 4. Modeling

We use the same Jupyter Notebook (`./lab.ipynb`) to analyze the data and model the distribution.

### 4.1. Pitch bend

How does microtonal pitch bend respond to pressure?

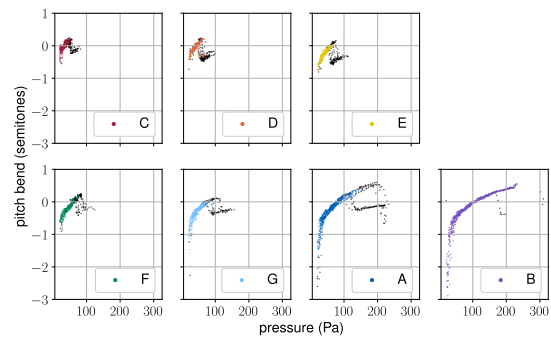

*Figure 5. Removal results.*
*The black dots correspond to discarded audio pages.*

Figure 5 shows a pitch bend scatter plot for each note. The non-equilibrium data discarded in section 3.4 are plotted as black dots. Clearly, the black dots don't belong to the main distributions.

Removing the black dots, overlaying all notes in one graph, and including the higher octave into the picture, we get Figure 6.

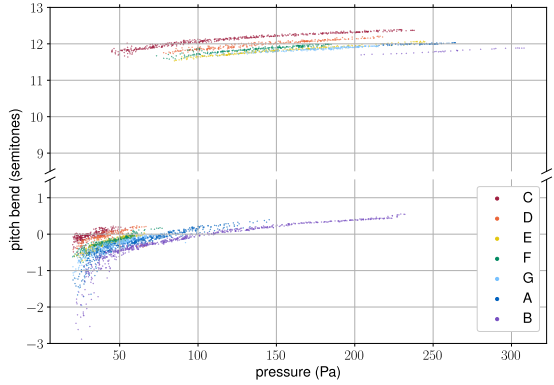

*Figure 6. Pitch bend against pressure.*
*The pressure is zeroed at atmospheric pressure.*

In Figure 6, the pitch bend increases as the pressure increases. The intersections with $y = 0$ and $y = 12$ correspond to "in-tune" moments. For a given color (i.e., pitch class), as the pressure rises, there is a sudden octave jump. The pressure threshold for this octave jump increases as the pitch class increases.

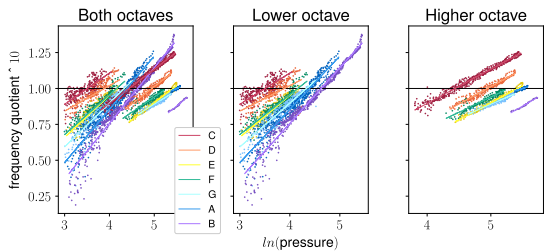

*Figure 7. A linear view on the pitch bend.*
*The fitted lines are overlaid.*

To fit a model, we transform the axes:
- The x axis (pressure) is now log scale.
- The y axis is changed from pitch bend into *frequency quotient* and raised to the $10^{th}$ power. The frequency quotient equals the actual frequency divided by the "in-tune" frequency.

That gives Figure 7, where we fit a linear model for each note (with the two octaves treated as different notes).

Next, we assume the slope of all lines to be the same and only the intercept may vary. That assumption gives new linear models with new intercept values. We plot the x-intercept (i.e., the $ln$ (pressure) that gives the "in-tune" frequency) against the pitch in Figure 8:

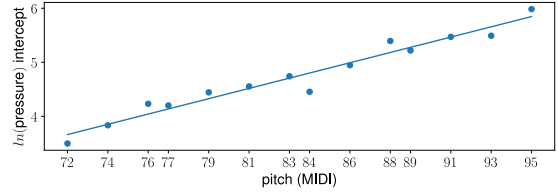

*Figure 8. X-intercept against pitch.*

We then fit a meta linear model for the (pitch, x-intercept) pairs. The modeling of pitch bend is now complete.

Note that there are two layers of parameter fitting here! To reiterate, from Figure 7 to Figure 8, the first round of fitting gives an array of linear models, and the second round of linear model fitting tries to explain the model parameters from the first round, which makes it meta.

The statistical complexity of this scheme is low. We only have three degrees of freedom: the "$10^{th}$" power applied to the frequency quotient, the slope, and the intercept of the meta linear model. With these three numbers[3], our model dictates that the pitch bend should respond to the pressure as depicted in Figure 9:

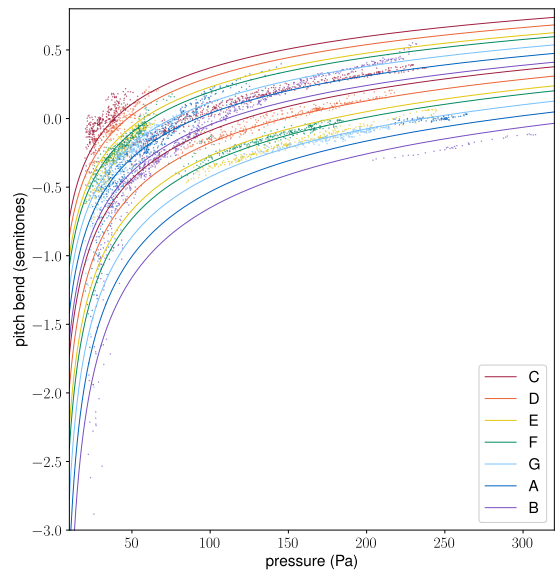

*Figure 9. Our pitch bend model, overlaid onto data.*

Figure 9 shows that our model, overall, agrees with our intuitions. However, it has poor accuracy with C6, E6, G6, and B6. Nevertheless, the flute that we measure can be imperfect. To us, the most problematic observation is that C6 is above B5 and it does not look like a flute imperfection. This undermines our assumption that the pitch bend behaviors only depend on the pitch – probably the flute being in the key of C breaks the symmetry and the flute "knows about" the octave difference between B5 and C6. Acoustic studies of wind instruments are likely to reveal the underlying mechanism and point out our mistakes in the model assumptions.

---

[3] The slope is $0.09498625513471028$. The intercept is $-3.177222804229106$.

## 4.2. Failure: amplitude study

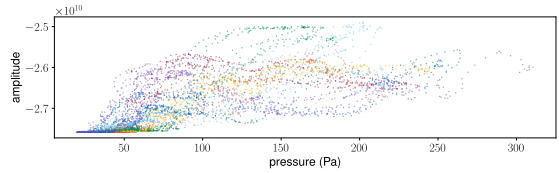

*Figure 10. Amplitude against pressure.*

As shown in Figure 10, we failed to find any relations between pressure and amplitude.

## 4.3. Octave hysteresis

Hysteresis is the phenomenon where an output variable prefers to stick to its previous state even when its input variable changes value for a small amount. Many wind instrument players are able to notice the octave hysteresis while playing. Octave hysteresis can be beneficial for octave stability but can become a challenge when one wants to perform rapid octave changes.

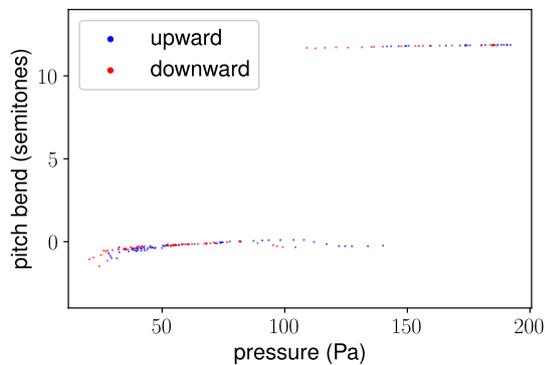

*Figure 11. Octave hysteresis.*

Figure 11 shows the pitch bend against the pressure for one note. The upward process (breath becoming stronger) is recorded by the blue dots, while the downward process (breath becoming weaker) is recorded by the red dots. The upward jump happens around 145 Pa while the downward jump happens around 105 Pa. Essentially, the octave hysteresis is captured in the difference between the pressure thresholds for up-octave and down-octave jumps. Therefore, we model it in section 4.4.

## 4.4. Pressure thresholds for octaves

We first label the pressure thresholds with a labeling tool (`./single_note_interactive_plot.py`). That gives the up-octave pressure threshold and the down-octave pressure threshold for each pitch class.

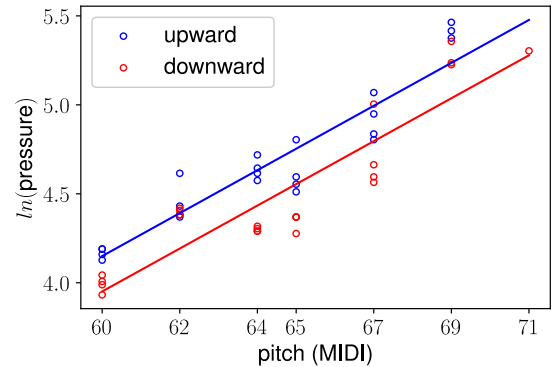

*Figure 12. Pressure thresholds for octaves, plotted for each pitch.*

In Figure 12, we have the log of pressure threshold value on the y axis. Each circle represents a datapoint, so generally there are four blue circles and four red circles for each pitch class, since each pitch class was played four times in section 2. The intercept difference between the blue (up-octave) line and the red (down-octave) line is calculated via adding the up-downness as a categorical input variable of the linear model.[4]

---

[4] The slope is 0.12067771159663639.
The blue intercept is −3.0908617599208004.
The red intercept is −3.289717945484731.

# 6. Appendices

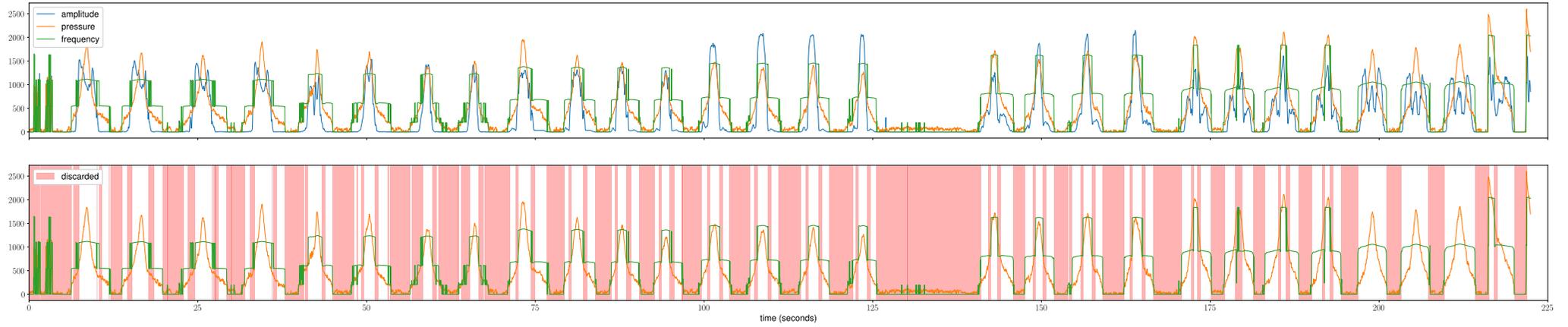

*Figure 13.*
*Upper half: overall time sequence of (breath) pressure, amplitude, and frequency.*
*Lower half: discarded data.*